\documentclass[11pt]{article}%
\usepackage[pdftex]{graphicx}
\usepackage{verbatim,color}
\usepackage{amsmath}
\usepackage{subfigure}
\usepackage{setspace}
\usepackage{subfig}
\usepackage{url}
\usepackage{float}

\usepackage[left=1in,top=1in,bottom=1in,right=1in, nohead]%
{geometry}
\usepackage{amsfonts}
\usepackage{amssymb}
\usepackage{graphicx}%
\begin{document}

\title{Sectoral co-movements in the Indian stock market: A mesoscopic network analysis}

\author{
Kiran Sharma\thanks{School of Computational and Integrative Sciences, Jawaharlal Nehru University, New Delhi-110067, India. Email: kiran34\_sit@jnu.ac.in}
\and Shreyansh Shah\thanks{Indian Institute of Technology, Banaras Hindu University, Varanasi-221005, India. 
Email: shreyansh.shah.mec13@iitbhu.ac.in}
\and
Anindya S. Chakrabarti\thanks {Economics area, Indian Institute of Management, Vastrapur, Ahmedabad, Gujarat-380015, India, Email: anindyac@iima.ac.in.} 
\and Anirban Chakraborti \thanks{School of Computational and Integrative Sciences, Jawaharlal Nehru University, New Delhi-110067, India, Email: anirban@jnu.ac.in}
}

\maketitle

\begin{abstract}
In this article we review several techniques 
to extract information from stock market data. We discuss recurrence analysis of time series, decomposition of
aggregate correlation matrices to study co-movements in financial data, stock level partial correlations with market indices,
multi-dimensional scaling and minimum spanning tree. 
We apply these techniques to daily return time series from the Indian stock market. 
The analysis allows us to construct networks based on correlation matrices of individual stocks in one hand and on the other,
we discuss dynamics of market indices. Thus both micro level and macro level dynamics can be analyzed using such tools.
We use the multi-dimensional scaling methods to visualize the sectoral structure of the stock market,
and analyze the comovements among the sectoral stocks. Finally, we construct a mesoscopic network based on sectoral
indices. Minimum spanning tree technique is seen to be extremely useful in order to separate technologically related
sectors and the mapping corresponds to actual production relationship to  a reasonable extent.  
\end{abstract}

\section{Introduction}
\label{intro}

In this paper, we present a coherent analysis of the Indian stock market employing several techniques
recently proposed in the econophysics literature. Stock market is a fascinating example of a rapidly evolving
multi-agent interacting system that generates an enormous amount of very well defined and well documented data.
Because of the sheer volume of data, it becomes possible to construct large scale correlation matrices across stocks
that contain information about the aggregate market. Thus the loss of information due to aggregation can be minimized to a great extent. Several useful techniques to analyze such large-scale data have been proposed and there are multiple
resources reviewing them. Interested readers can refer to \cite{Mantegna_Stanley} and \cite{Bouchaud_Potters}
for excellent and quite extensive textbook expositions.

We present a series of analysis on the Bombay stock exchange, using both macro scale and micro scale data.
Even though there are separate attempts in a few other papers that presented analysis on similar data sets, this probably
is the first attempt to systematically analyze Indian stock market data in a comprehensive manner.
At the beginning of discussion on every technique, we point out the papers that proposed the techniques and
subsequent analysis, if any, on Indian or any other emerging market data. 
India being an emerging market is an interesting example.
Several papers (Ref. \cite{RP_Bastos}, \cite{Sinha_Pan;08}) have pointed out that
there are systematic differences between the dynamic behavior of developed economies and emerging economies.

\section{Nonlinear dynamics: recurrence plot analysis}
\label{nonlinear}

For a very long time it had been conjectured that the stock market indices may have certain features of
a highly
nonlinear dynamical system. 
It originated from certain speculations that economic systems in general may show chaotic behavior 
(see e.g. \cite{Baumol_Benhabib_89}). \cite{Brock_Sayers_88} considered an idea that the
aggregate macro dynamics of an economy may show chaotic behavior. By and large, such thoeries 
are no longer considered to be useful descriptions of economic dynamics. However, in recent times
there have been some attempts to analyze the stock index behavior by using recurrence analysis
based on phase space reconstruction.

\begin{figure}
	\centering
		\includegraphics[width=0.7\linewidth]{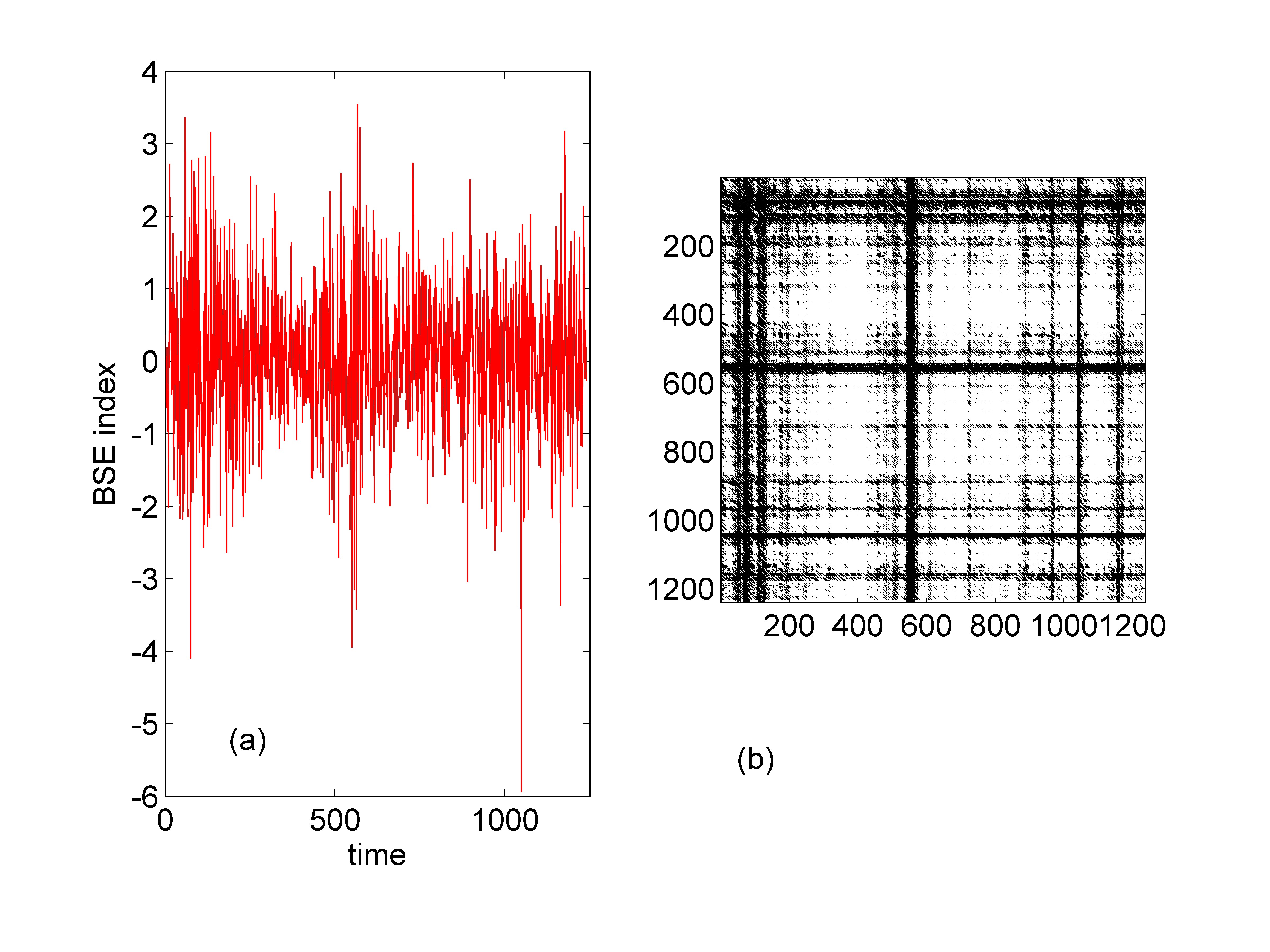}
\caption{
Left panel: Normalized daily return series constructed from BSE index data for five years (6th June, 2011 to 6th June, 2016).
Right panel: Recurrence plot constructed from the same data with an embedding dimension equals to 11 and time delay 1.
}
	\label{Fig:RP_1}
\end{figure}

In general, the technique's usefulness comes from the fact that it is non-parametric, does not make any assumptions
about the data and can work with non-stationary data. In particular, the technique is useful for detecting sudden large change in a time series. A stock market crash has often been thought of as a phase transition
indicating a large abrupt change in the behavior \cite{Sornette_book}.
However, the technique is useful for recovering
patterns in potentially highly nonlinear but recursive systems, an assumption that is not satisfied by the stock market.
We follow the mode of analysis presented in details in \cite{RP_Guhathakurta} and \cite{RP_Bastos}.

\begin{figure}
	\centering
		\includegraphics[width=0.7\linewidth]{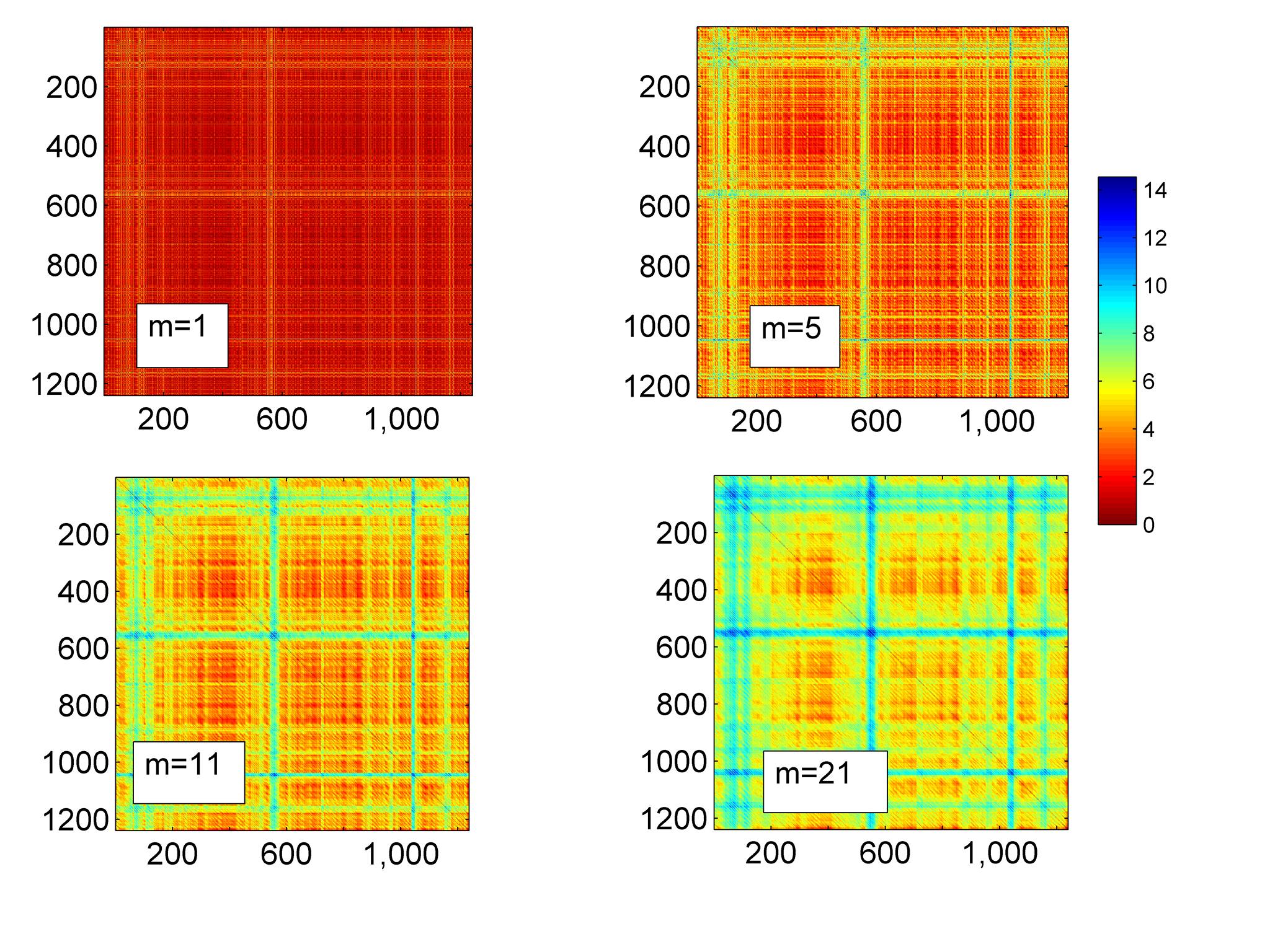}
	\caption{
Distance plots constructed from the BSE index for different values of the embedding dimensions ($m=~ 2,~5,~11,~21$).
}
	\label{Fig:RP_2}
\end{figure}

Here we describe construction of recurrence plots. 
It is based on the idea of recurrence within a phase space and the plot exhibits times when a nonlinear system revisits the same phase space during the process of evolution.
Consider a time series $\{x(i)\}_{i=1}^{N}$ representing an index of a stock market. We know from Takens' theorem \cite{Takens_81} that it is possible to extract information about the phase space from the time series (see also Ref. \cite{RP_Bastos}). 
We start by embedding $\{x\}$ into an $m$ dimensional space given by,
\begin{equation}
y(i) =   [x(i), x(i+\delta), x(i+2\delta),..., x(i+(m-1)\delta)]
\end{equation}
where, $d$ is the time delay. Together these two parameters constitute the set of embedding parameters. 
Thus $y(i)$ is a point in the $m$ dimensional Euclidean space, representing the evolution of the system in the reconstructed phase space.
We collect all such $y(i)$'s and present element-by-element difference with Euclidean norm to create a two dimensional plot. Such a plot exhibits
if there is any recurrence as explained below.

Let us define a matrix $R$ such that its $i,j$-th element ($i,j = 1,....,n,$ with $n = N - (m-1)d$) is expressed as
\[
    R_{ij}(\epsilon)=\left\{
                \begin{array}{ll}
                  ~0~~\mbox{if~~} |y(i) - y(j)| > \epsilon \\
                  ~1 ~~ \mbox{if~~} |y(i) - y(j)| \leq \epsilon
                \end{array}
              \right.
\]
where $\|.\|$ is the Euclidean norm, and $\epsilon$ is the threshold applied which is a positive real number.
Recurrence plots are exactly symmetric along the diagonal.

\textbf{Inference based on structures}:

In recurrence plots, we see multiple patterns including dots, diagonal as well as vertical and horizontal lines and all possible combinations of them.

\begin{itemize}
	\item Isolated points exist if states are rare or persistence is low or if they represent high fluctuations. 
	\item Existence of a diagonal line $R_{i+m, j+m}$ = 1 (for $m=1,\ldots,l$ where $l$ is the length of the diagonal line) indicates presence of
recurrence i.e. a segment of the time series revisits the same area in the phase space at a lag.
If there are lines parallel to the line of identity, it represents the parallel evolution of trajectories. 
	\item Existence of a vertical (horizontal) line $R_{i,j+m}$ = 1 (for $m=1,\ldots,v$ where $v$ is the length of the line) 
indicates a stage during evolution where the system gets trapped for some time and does not evolve fast. This can be an intermittent behavior.
\end{itemize}

Now we conduct recurrence quantification analysis (RQA) by studying the structure of the plots numerically. Such an analysis is based essentially on densities of isolated points,
diagonal lines as well as vertical lines. We borrow the discussion presented below from \cite{RP_Bastos}. The measures which we have considered are as follows:
\begin{itemize}
	\item RR: Recurrence rate.
           \item DET: Fraction of points in the plot forming diagonal lines. This indicates determinism and hence, predictability.
	\item $\langle L \rangle$: Average lengths of the diagonal lines.
	\item LMAX: Length of the longest diagonal line (except the line of identity). Its inverse is associated with the divergence of the trajectory in phase-space.
	\item ENTR: Shannon entropy defined over the distribution of lengths of diagonal lines, indicates diversity of the diagonal lines.
	\item LAM: Fraction of points forming vertical lines, indicates existence of laminar states in the system.
	\item TT: Average length of the vertical lines, this value estimates the trapping time.
\end{itemize}
We have computed the RQA measures for the BSE index under a range of embedding dimension. In each case, we have set the delay equal to 1.
\begin{table}
	\centering
		\begin{tabular}{ |c|c|c|c|c|}
                     \hline
                     Quantity & $m=1$ & $ m=2$ & $m=5$ & $m=11$ \\
		\hline\hline
                      RR & 0.0758 & 0.0442 & 0.0075 & 3.1049$\times~10^{-4}$ \\
                      \hline 
		DET & 0.9029 & 0.8817 & 0.8516 & 0.9211\\
		\hline
		$\langle L\rangle$ & 4.3854 & 3.9302 & 3.8841 & 4.6667 \\
		\hline
		LMAX & 146 & 88 & 58 & 18 \\
		\hline
		ENTR & 2.0319 & 1.8575 & 1.8095 & 1.8527 \\
		\hline
		LAM & 0.9479 & 0.9074 & 0.7234 & 0.2763  \\
		\hline
		TT & 5.7416 & 4.4451 & 3.2874 & 2.2703 \\
		\hline
		\end{tabular}
\caption{Measures based on recurrence analysis of normalized BSE data. Generated by the CRP toolbox (Ref. \cite{CRPtoolbox} and \cite{CRPtoolbox_paper}). Threshold for calculating neighbors set at the default value 0.1.}
	
\label{table:RQA}
\end{table}
In Fig. \ref{Fig:RP_1} and \ref{Fig:RP_2} we present recurrence analysis on logarithmic return series ($r_{\tau } = \ln P_{\tau}- \ln P_{ \tau -1}$) constructed from BSE index data. As is apparent, there is no clearly discernible pattern in the data.
Next, we follow the standard approach and use the level of the price data upon normalization by the maximum value of the time series ($\tilde{P}_{\tau}=(P_{\tau}-P_{min})/P_{max}$).
Table \ref{table:RQA} contains the RQA measures. Most of the prior literature on stock market data consider high values of embedding parameter. It is evident that, in general, recurrence rates are very low and determinism is very high. 
However, this approach has an inherent problem that it is not particularly good at differentiating non-recursive series from recursive series.
In general, we found that when we construct similar measures for standard recursive series, it is not clear from such RQA measures that they can be easily separated from a stochastic series. Thus it does not really shed much light on the problem as the primary focus is to figure out determinism or lack thereof. Ref. \cite{RP_Bastos}
discusses a possible application that these measures still retain some usefulness for cross-country analysis. Since in this case we are focusing on one country only,
it is not very helpful.
So we consider a fully stochastic framework in the rest
of the analysis.

\section{Empirical study of the correlation structure of the Indian stock market}
\label{emp_study}

In this section, we analyze the empirical cross-correlation matrices constructed from the stock  market data.

\subsection{Data specification, notations, and definitions} 

In order to study correlations and co-movements in the stock price time series, the popular Pearson correlation coefficient was commonly used. However, with the electronic markets producing data at different frequencies (low to high), it is now known that several factors viz., the statistical uncertainty associated with the finite-size time series, heterogeneity of stocks, heterogeneity of the average inter-transaction times, and asynchronicity of the transactions may affect the applicability/reliability of this estimator. In this article, we have mainly focused on the daily returns computed from closure prices, for which the Pearson coefficient works well.

\subsubsection{Data set}
\label{data}

We have used the freely download able daily adjusted closure prices from Yahoo finance  for $ N = 199$ companies in the Bombay Stock Exchange (BSE) SENSEX \cite{Yahoo}, for five years, over a period spanning from June 6, 2011 to June 6, 2016. Also we have downloaded 199 stock prices of companies chosen randomly from the BSE and 13 sectoral indices of the BSE, for the period  May 27, 2011 to May 27, 2016. The lists are given in  the appendices I and II.

\begin{figure}
	\centering
		\includegraphics[width=0.7\linewidth]{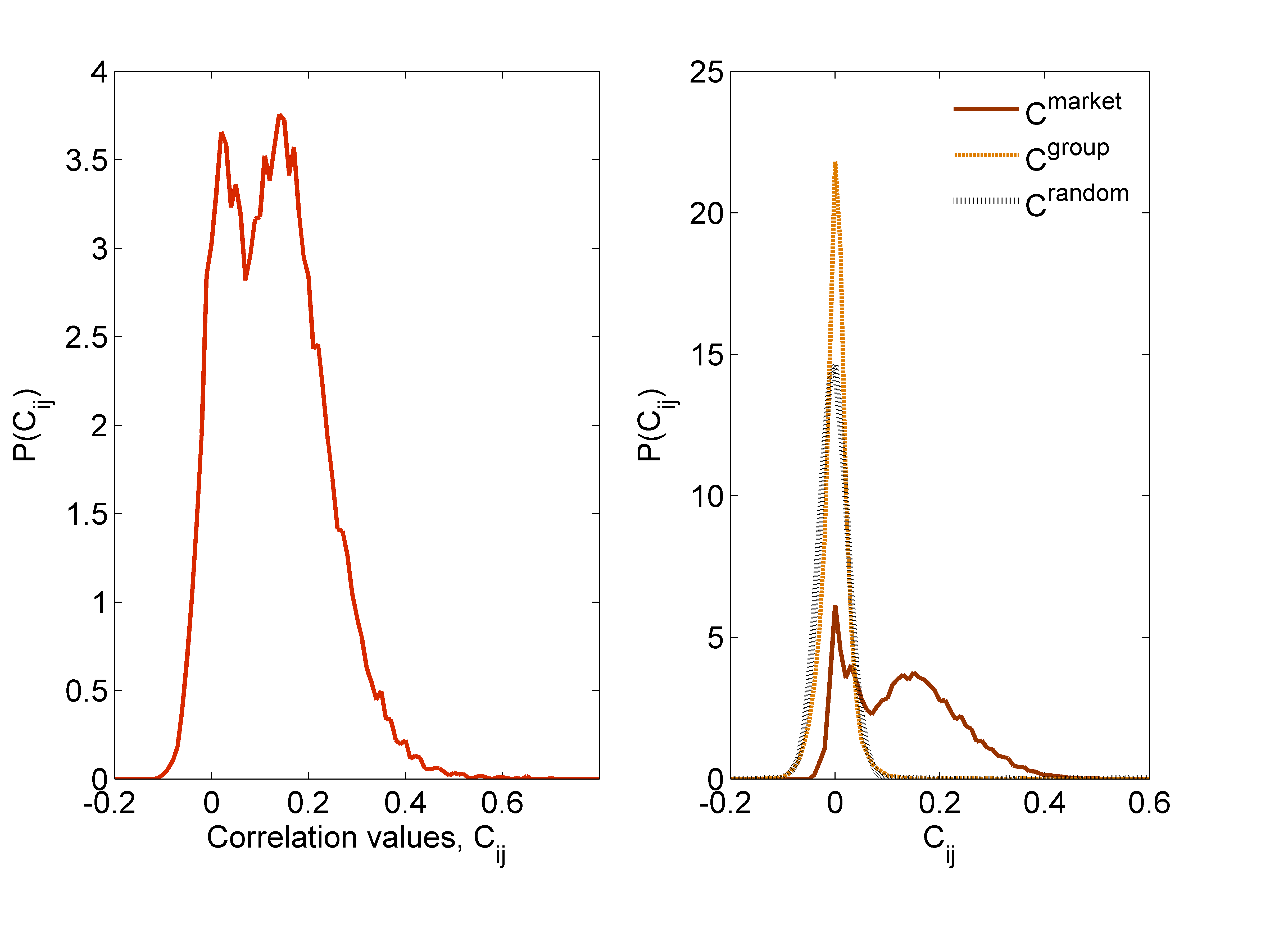}
\caption{
Left panel: Probability density function of the cross-correlation coefficients of 199 BSE stocks.
Right panel: Decomposition of the correlation matrix into market mode, group mode and random mode.
}
	\label{Fig:corr}
\end{figure}

\subsection{Correlation matrices}
\label{correlation}

We construct the correlation matrix from individual stock returns in the following way.
\subsubsection{Pearson correlation coefficient}

In order to study the equal time cross-correlations between $N$ stocks, we first denote the adjusted closure price of stock $i$ in day $\tau$ by $P_i(\tau )$, and determine the logarithmic return of stock $i$ as $r_i( \tau ) = \ln P_i( \tau )- \ln P_i( \tau -1)$. For the window of $T$ consecutive trading days, these returns form the return vector $r_i$. We use the equal time Pearson correlation coefficients between stocks $i$ and $j$ defined as
\begin{eqnarray}
C_{ij} = \frac{\langle r_i r_j \rangle - \langle r_i \rangle \langle r_j \rangle}{\sqrt{ [\langle r^{ 2}_i \rangle - \langle r_i \rangle^2][\langle r^{ 2}_j \rangle - \langle r_j \rangle^2]}}, 
\label{Eq:pearson_corr}
\end{eqnarray}
where $\langle...\rangle$ indicates an average over the window of  $T$ successive trading days in the return series. Naturally, such correlation coefficients satisfy the usual condition of $−1\leq C_{ij} \leq1$ and we can create an $N \times N$ correlation matrix $C$ by collecting all values \cite{Chakraborti;06,Tilak;12}. By construction, the matrix is symmetric and
it serves  as the basis of the rest of the present article. 

\begin{figure}
	\centering
		\includegraphics[width=0.7\linewidth]{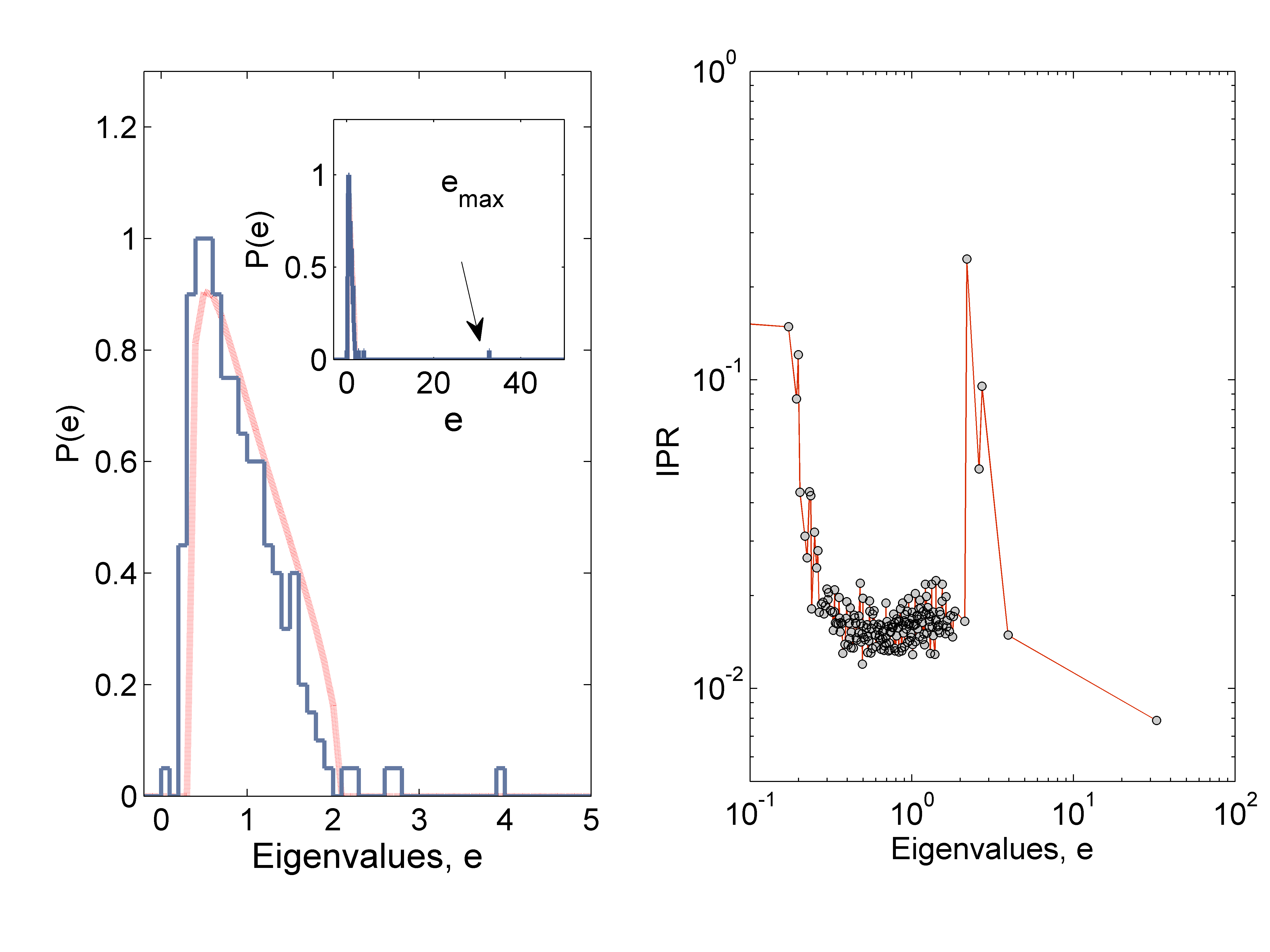}
	\caption{
Eigenvalue decomposition of the correlation matrix. Left panel: Probability density function of eigenvalues. Inset shows the full distribution.
Right panel: Inverse participation ratio with respect to the corresponding eigenvalues.
}
	\label{Fig:eig_val}
\end{figure}

\subsection{Decomposition analysis}
\label{corr_decomp}

For the present section, we are following the sequence of methods discussed by Ref. \cite{Sinha_Pan;08} which is one of the first few papers that
applied this technique.
Suppose we have $N$ return time series of length $T$ that are pairwise uncorrelated. The correlation matrix generated by collecting all pair-wise correlations for $N$ such series is called Wishart matrix. In the limit $N \rightarrow \infty$ and $T \rightarrow \infty$, such that the ratio $Q \equiv T /N > 1$, the eigenvalue distribution of this matrix has a specific distributional form,
\begin{equation}
f(\lambda) = (Q/2\pi)\frac{\sqrt{(\lambda_{max} - \lambda)(\lambda - \lambda_{min})}}{\lambda},
\end{equation}
for $\lambda_{min} \leq \lambda \leq \lambda_{max}$ and, 0 otherwise. This distribution is clearly bounded by $\lambda_{max,min} = [1 \pm (1/\sqrt{Q})]^2$. In the BSE data we considered, $Q$ = 5. Thus the Wishart matrix should have the following bounds: $\lambda_{min}$ = 0.3056 and $\lambda_{max}$ = 2.0944. The distribution of eigenvalues
unexplained by the Wishart matrix  sheds light on the interaction structures and the coevolution process of the stocks in the market.

The largest eigenvalue corresponds to the market mode which captures the aggregate dynamics of the market that is common across all stocks. 
The eigenvectors associated with the next few eigenvalues (we took the next 5 dominant eigenvalues) describe the sectoral dynamics.
The rest of the eigenvectors correspond to the random mode. 
From such a segregation, it is possible to reconstruct the contributions of different modes to the aggregate correlation matrix.

Following the literature to filter the data to remove market mode and the random noise, we first decompose the aggregate correlation matrix as
\begin{equation}
C = \sum_{i=0}^{N-1} \lambda_{i}a_{i}a_{i}^{T} ,
\end{equation}
where $\lambda_{i}$ are the eigenvalues of the correlation matrix $C$. An easy way to handle the reconstruction of the correlation matrix is to sort the eigenvalues in descending order. Then we rearrange the eigenvectors $a_{i}$ in corresponding ranks. 
This allows us the decompose the matrix into three separate components viz. market, group and random:
\begin{eqnarray}
C &=& C^{M} + C^{G} + C^{R},\\
   &=& \lambda_{0}a_{0}a_{0}^{T} + \sum_{i=1}^{N_{G}} \lambda_{i}a_{i}a_{i}^{T} + \sum_{i=N_{G}+1}^{N-1} \lambda_{i}a_{i}a_{i}^{T}
\end{eqnarray}
where $N_{G}$ is taken to be 5 i.e. corresponds to the 5 largest eigenvalues except the first one. 
It is worth noting that the exact value of $N_G$  is not crucial for the result as long as it is kept within the same ballpark.
The decomposition is shown in Fig. \ref{Fig:corr}.

An important finding is that the group mode almost coincides with the random mode whereas the market mode is segregated by a large margin from the rest.
Thus the sectoral dynamics are almost absent whereas the market mode is very strong. This is in line with the prior literature (see e.g. \cite{Sinha_Pan;08}).

Following standard procedure (see e.g. Ref. \cite{Sinha_Pan;08}), we also calculate the inverse participation ratio (IPR)
to extract information about contribution of different stocks to the eigenvalues.
IPR is defined for the $k$-th eigenvector as the sum of fourth power of all individual components of the corresponding eigenvector, $I_{k} \equiv \sum_{i=1}^{N} [a_{ki}]^4$, where $a_{ki}$ are the components of eigenvector k. 
The result is presented in Fig. \ref{Fig:eig_val}. 
Intuitively  if a single stock dominates in terms of contribution to any particular eigenvector, 
then the $IPR$ would go to 1. For example, consider a limiting case of
$a_{k1} = 1$ and $a_{ki} = 0$ for $i \neq 1$.
On the other hand if all elements were equal to $1/\sqrt{N}$, then we would get $IPR=1/N$.
Thus by considering IPR, we can understand if there is significant contribution coming from specific stocks or a more diversified bundle of stocks.

\subsection{Partial correlation analysis}
\label{partial_correlation}

Partial correlation is a newly introduced tool to investigate the effects of third stock on the correlation between a pair of stocks. 
\cite{Kenett_15} introduced this analysis for multiple stock markets. In the present paper we apply their technique to the Bombay stock exchange data.
To describe its usefulness, consider 3 stocks, $i$, $j$ and $k$, with significant correlation between all three pairs of the stocks.
Suppose, we think that the high value of $C_{ij}$ is the result of their own correlations with $k$, i.e. part of $C_{ij}$ might
be spurious correlation arising from a third variable effect (in this case $k$),
we should remove such effects to figure out the actual correlation across $i$ and $j$.
Then we can recalculate $C_{ij}$, after controlling for the effect of $k$. 
The resultant correlation value is called the partial correlation.
The difference between the raw correlation value between a pair of stocks and the corresponding partial correlation tells us how much third-variable effect was there.

\begin{figure}
	\centering
		\includegraphics[width=0.7\linewidth]{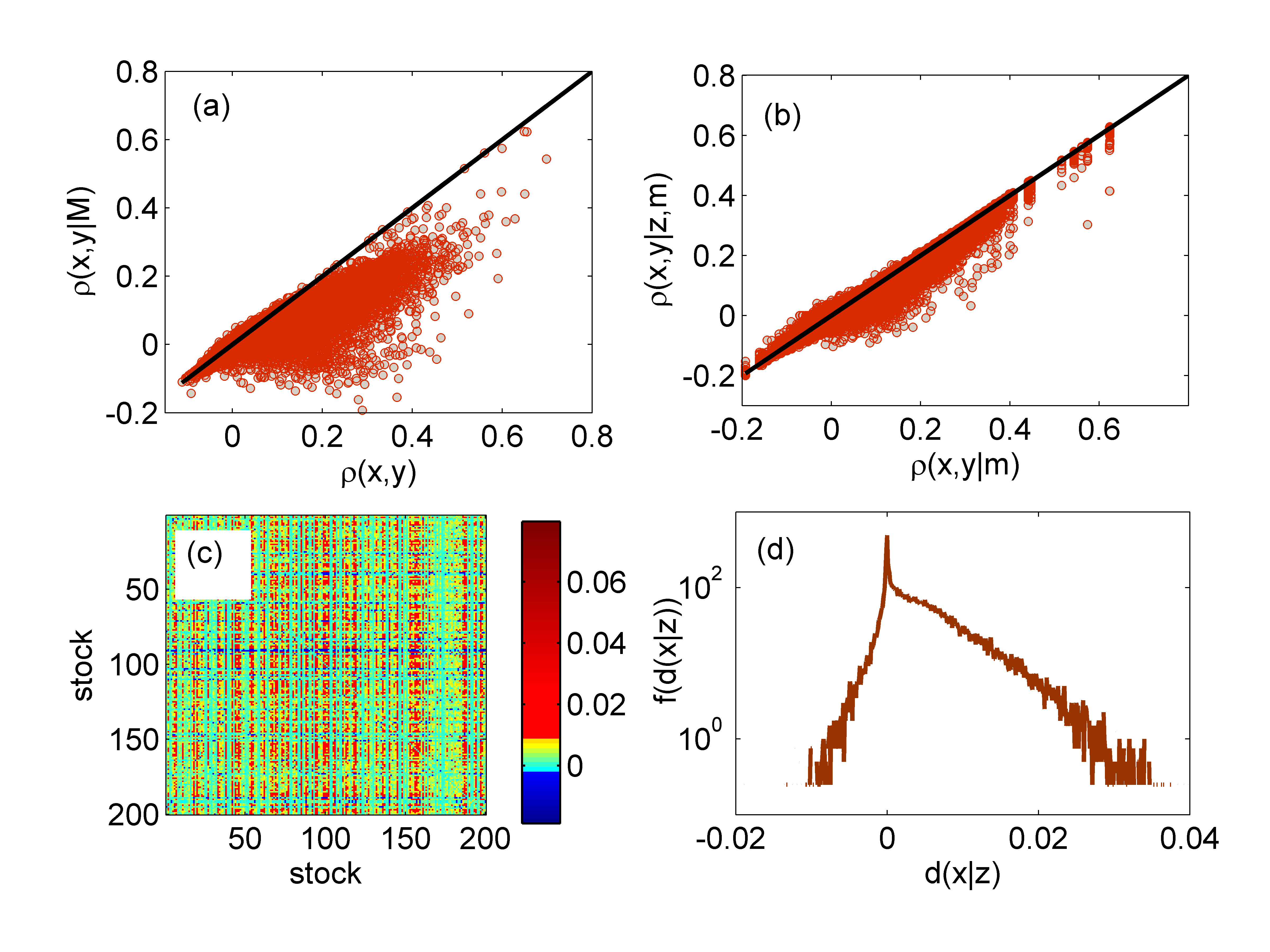}
	\caption{
Correlation matrix after controlling for market mode (BSE index). Panel (a): Partial correlation after controlling for the market mode as a function of raw correlation coefficients.
Panel (b): Same after controlling for the third variable effect. Panel (c): Influence of all stocks as the third variable (on $x$-axis) on all other stocks (on $y$-axis) . Panel (d): Probability density function of average influence quantity.
}
	\label{Fig:rho_xym}
\end{figure}

For this purpose, we again use the same daily log return $r_{i}(t)$. 
However, we need to adjust for one more factor. From the preceding analysis, we already know that there is a significant market mode.
Therefore, that will act a common driving factor. Hence, the market mode should also be controlled for in order to extract the actual correlation values for the exact same reason.
In this case, the market mode is given by a market index. 
Note the difference from the earlier analysis. For constructing the market mode from the eigenvalue analysis, the market mode arise endogenously from the
panel data itself whereas in this case, we take the market mode to be given by an exogenous index time series. Hence, these two types of analysis complement each other.

Following the notation of \cite{Kenett_15}, let $x$, $y$ be two time series and let $M$  be the BSE index for the same time frame. The partial correlation, $(x,y| M)$
is defined as the standard Pearson correlation coefficient (described above) between $x$ and $y$ after controlling for $M$. More technically, this is the correlation between the
residuals of $x$ and $y$ which are unexplained by the market index represented by $M$. So first, we need the residuals of the two time series. A simple way is to do it would be to regress
both on M. Then we can work with the resulting variables. Formally, the correlation is given as 
\begin{equation}
C_{x,y|M} = \frac{(C_{x,y} - C_{x,M}.C_{y,M})}{\sqrt{[1-C_{x,M}^2].[1-C_{y,M}^2]}}
\end{equation}

In the same way, when the same two stocks $x$ and $y$ are affected by a common stock $z$, 
we can control for that effect as well. Given a third stock $z$, the partial correlation between $x$ and $y$ after controlling for both the market factor as well as that third stock $z$, is
given by the following formula 
\begin{equation}
C_{x,y| M, z} =\frac{C_{x,y|M}- C_{x,z|M}.C_{y,z|M}}{\sqrt{[1-C_{x,z|M}^2].[1-C_{y,z|M}^2]}}
\end{equation}

If it is found that the third stock has an important effect on pairs of stocks, then it is useful to define the `Influence Quantity' (see \cite{Kenett_15})
\begin{equation}
d(x,y|z) = C_{x,y|M}- C_{x,y|M,z}.
\end{equation}
Magnitude of this quantity will reflect how much influence does the third stock influence on a pair of stocks. 
A natural extension of this idea is to consider the average influence $d(x|z)$ of stock $z$ on the correlations between a given stock $x$ and 
all other stocks except $x$ itself and $z$. Ref.
\cite{Kenett_15} defined this index as the following
\begin{equation}
d(x|z) = \left\langle d(x,y|z) \right\rangle_{y \neq x}.
\end{equation}
This quantity captures the average influence from stock $z$ to stock $x$ through the third variable effect after controlling for the market index.

We present all results of our analysis in Fig. \ref{Fig:rho_xym}. Panel (a) shows the correlation coefficients of all stocks after controlling for the market index.
Since the bulk of it is below the 45$^{\circ}$ line, we conclude that the market index has a positive effect on pairwise correlations. This is consistent with the results from the
eigenvalue analysis and also with \cite{Kenett_15}. Similarly, in panel (b) we show the data for the same correlation coefficients after controlling for all possible third variable effects.
In panel (d), we present the probability density function of the {\it influence quantity}. Again bulk of the distribution is in the positive quadrant implying positive effect on average.

\section{Network analysis}
In this section, we present network analysis based on the empirical correlation matrix.
\subsection{ Distance metric}

To obtain ``distances", the following transformation
\begin{eqnarray}
 d_{ij} = \sqrt{2 (1- C_{ij})},
 \label{Eq:distance}
\end{eqnarray}
is used, which clearly satisfies $2 \geq d_{ij} \geq 0$. Collecting all distances one can form an $N \times N$ distance matrix $D$, such that all elements of the matrix are ``ultrametric" \cite{Rammal;86}. The concept of ultrametricity appears in multiple papers. Interested readers can refer to the detailed discussions by Mantegna \cite{Mantegna;99,OnnelaI;03,OnnelaII;03,Chakraborti;06} among others. There are multiple possible ultrametric spaces. We opt for the subdominant ultrametric, as it is simple to work with and and its associated topological properties. The choice of the non-linear function is again arbitrary, as long as all the conditions of ultrametricity are met.

\subsection{Multidimensional scaling (MDS)} 
\label{mds}

Multidimensional scaling is a method to analyze large scale data that displays the structure of similarity in terms of distances, given by Eq. \ref{Eq:distance}, as a geometrical picture or map, where each stock corresponds to a set of coordinates in a multidimensional space. MDS arranges different stocks in this space according to the strength of the pairwise distances between stocks, -- two similar stocks are represented by two set of coordinates that are close to each other, and two stocks
behaving differently are placed far apart (see Ref. \cite{Borg;05}) in the space. We construct a distance matrix consisting of $N \times N$ entries from the $N$ time series available, defined using Eq. \ref{Eq:distance}:
\begin{eqnarray}
D = 
\begin{bmatrix}
d_{11} & d_{12}&\ldots & d_{1N} \\
d_{21} & d_{22} &\ldots & d_{21} \\
\vdots & \vdots & \ddots & \vdots\\ 
d_{N1} & d_{N2} &\ldots & d_{NN} \\
\end{bmatrix}.
\end{eqnarray}

Given $D$, the aim of MDS is to generate $N$ vectors $x_1,...,x_N \in \Re^D$, such that
\begin{eqnarray}
\Arrowvert x_i  -  x_j \Arrowvert \approx d_{ij} \hspace{0.3in}\forall i, j\in N, 
\end{eqnarray}
where $\Arrowvert . \Arrowvert$ represents vector norm. We can use the Euclidean distance metric as is done in the classical MDS. Effectively, through MDS we try to  find a mathematical embedding of the $N$ objects into $\Re^D$ by preserving distances. In  general, we choose the embedding dimension $D$ to be  $2$, so that we are able to plot the vectors $x_i$ in the form of a map representing $N$ stocks. It may be noted that $x_i$ are not necessarily unique under the assumption of the Euclidean metric, as we can arbitrarily translate and rotate them, as long as such transformations leave the distances $\Arrowvert x_i - x_j \Arrowvert$ unaffected. Generally, MDS can be obtained through an optimization problem, where $(x_1,...,x_N)$  is the solution of the problem of minimization of a cost function, such as
\begin{eqnarray}
\min_{x_1,...,x_N} \sum_{i<j} (\Arrowvert x_i - x_j \Arrowvert - d_{ij})^2.
\end{eqnarray}

\begin{figure}[H]
	\centering
		\includegraphics[width=0.7\linewidth]{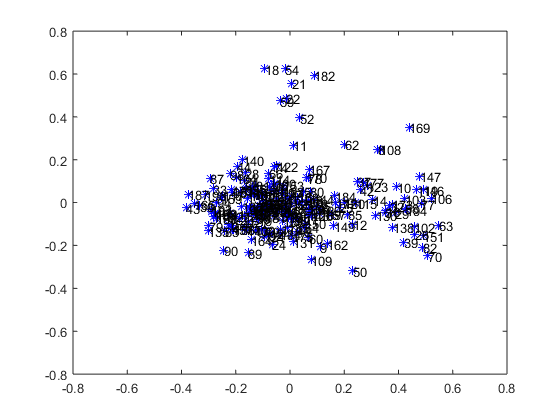}
	\caption{
Multidimensional scaling of the sample data for the time window May 2015-May 2016. }
	\label{Fig:MDS}
\end{figure}

In order to capture the sectoral behavior of the market visually, we have generated the MDS plot of 199 stocks as described before, for the time window of 250 trading days between May 2015 - May 2016. As before, using the correlation matrix as input, we computed the distance matrix using the transformations (given by Eq. \ref{Eq:distance}). The distance matrix was then used as an input to the inbuilt MDS function in MATLAB \cite{mds}. The output of the MDS were the sets of coordinates, which were plotted as the MDS map as shown in Fig. \ref{Fig:MDS}. 

The coordinates are plotted in a manner such that the centroid of the map coincides with the origin $(0,0)$. It is interesting to follow the positions of certain sectors: (i) Sugar, (ii) Textiles and (iii) Pharmaceuticals, which will be discussed in details in Section \ref{sectoral}.

\subsection{Dendrogram }
\label{dendro}

\begin{figure}[H]
	\centering
		\includegraphics[width=0.7\linewidth]{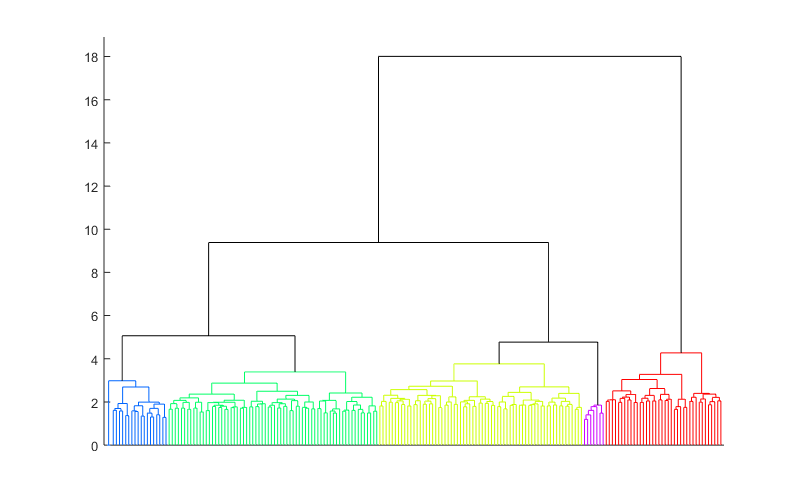}
	\caption{
Dendrogram of 199 stocks.
}
	\label{Fig:Dendrogram}
\end{figure}

Dendrogram is basically a tree diagram. This is often used to depict the arrangement of multiple nodes through hierarchical clustering. We have used the inbuilt function in MATLAB \cite{dendrogram} to generate the hierarchical binary cluster tree (dendrogram) of $N$ stocks connected by many U-shaped lines (as shown in Fig. \ref{Fig:Dendrogram}), such that the height of each U represents the distance (given by Eq. \ref{Eq:distance}) between the two data points being connected. Thus, the vertical axis of the tree captures the similarity between different clusters whereas the horizontal axis represents the identity of the objects and clusters. Each joining (fusion) of two clusters is represented on the graph by the splitting of a vertical line into two vertical lines. The vertical position of the split, shown by the short horizontal bar, gives the distance (similarity) between the two clusters. We set the property ``Linkage Type" as ``Ward’s Minimum Variance", which requires the Distance Method to be Euclidean which results in group formation such that the pooled within-group sum of squares would be minimized. In other words, at every iteration, two clusters in the tree are connected such that it results in the least possible increment in the relevant quantity i.e. pooled within-group sum of squares. Fig. \ref{Fig:Dendrogram} shows the dendrogram of all the 199 stocks clustered in five different colors (by using `ColorThreshold' property in MATLAB). The magenta color represents the cluster of `Sugar Industries'.

\subsection{Minimum spanning tree}
\label{mst}
A minimum spanning tree is a spanning tree of a connected, undirected graph such that all the $N$ vertices are connected together with the minimal total weighting for its $N-1$ edges (total distance is minimum).  The distance matrix defined by Eq. \ref{Eq:distance} was used as an input to the inbuilt MST function in MATLAB \cite{mst}. See Matlab documentation for all details. Here we state Kruskal and Prim algorithms for the sake of completeness of the present article. 
Description of the two algorithms (source: see Ref. \cite{mst}):
\begin{itemize}
\item Kruskal --  {\it This algorithm extends the minimum spanning tree by one edge at every discrete time interval by finding an edge which links two separate trees in a spreading forest of growing minimum spanning trees}. 
\item Prim -- {\it This algorithm extends the minimum spanning tree by one edge at every discrete time interval by adding a minimal edge which links a node in the growing minimum spanning tree with one other remaining node}.
\end{itemize}

\begin{figure}[H]
	\centering
		\includegraphics[width=0.7\linewidth]{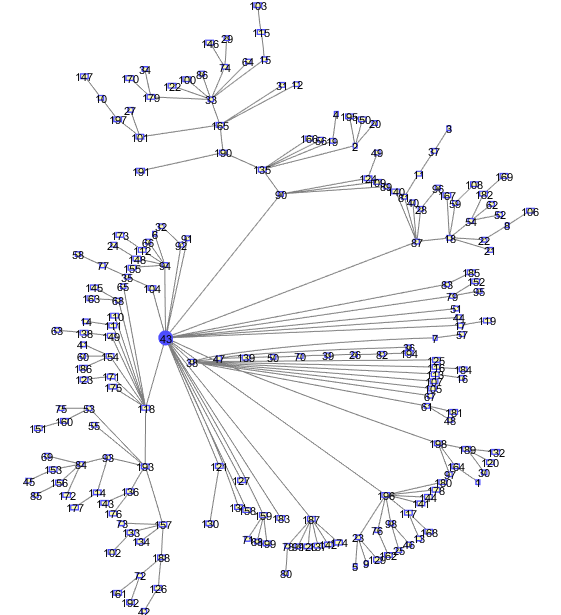}
	\caption{
Minimum spanning tree of the sample data.
}
	\label{Fig:MST}
\end{figure}

Fig. \ref{Fig:MST} shows the MST for all the 199 stocks. Matlab algorithms set the root node as the first node in the largest connected component, which in our case is node 43.

\section{Sectoral co-movement: mesoscopic network}
\label{sectoral}


After quantifying the general cross-correlation structure of the market, we probe deeper into the sectoral
co-movements. There are multiple ways to analyze the data. One, we can impose a threshold on the
group cross-correlation matrix and construct a network of stocks which move closely. This is the approach that is
followed in \cite{Sinha_Pan;08} for example.
This approach has some problems. One, the threshold has to be exogenous and hence, basically arbitrary.
Two, even with such networks, it is difficult to identify clusters that matches with actual industry classifications.
An alternative way is to follow the industry classifications first and then try to see if they form 
clusters.

\begin{figure}
	\centering	
\includegraphics[width=0.45\linewidth]{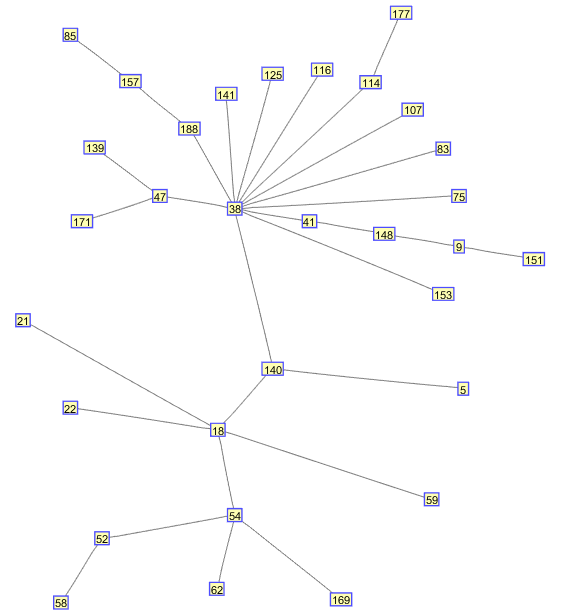}
	\caption{
Plot of MST for the sectors.
}
	\label{Fig:sector_plots}
\end{figure}

To study the sectoral behavior in the market, we have selected stocks from the list of BSE from the industries: (i) Sugar, (ii) Textiles and (iii) Pharmaceuticals. Following the same methodologies as described in the previous sub-sections \ref{mds}, \ref{dendro} and \ref{mst}, we have generated the plots given in Fig. \ref{Fig:sector_plots}.

By looking at the diagram, it becomes clear that the method is partially successful to segregate the market into clusters,
but not fully. Therefore, we construct a new network. Rather than working with actual stock returns, we
work with sectoral index returns. 

This marks a prominent departure from the usual mode of analysis. Typically, most studies focus on
either an aggregate macro-level market index like S\&P 500, or consider collective dynamics of
micro-level individual stock returns. here we consider a  {\it mesoscopic} network to characterize correlations.

\begin{figure}
	\centering
		\includegraphics[width=0.6\linewidth]{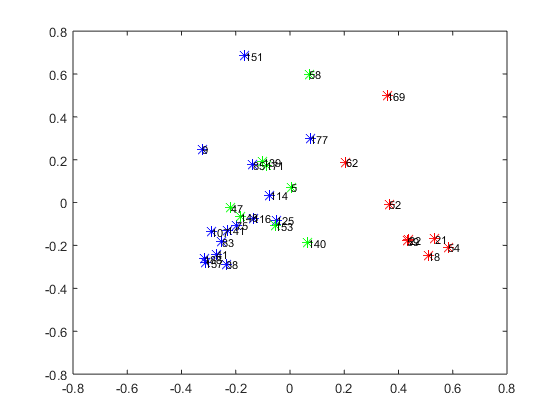}
\caption{Plot of MDS for the sectors: Sugar (red), Pharmaceuticals (green) and Textiles (blue).}
\label{Fig:MDS_sector}
\end{figure}

\begin{figure}
	\centering	
\includegraphics[width=0.6\linewidth]{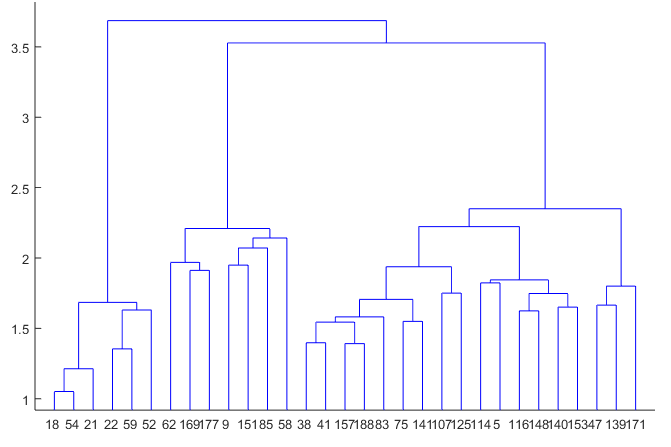}
\caption{Plot of dendogram for the sectors.}
\label{Fig:dendogram_sectors}
\end{figure}

\begin{figure}[H]
	\centering
		\includegraphics[width=0.5\linewidth]{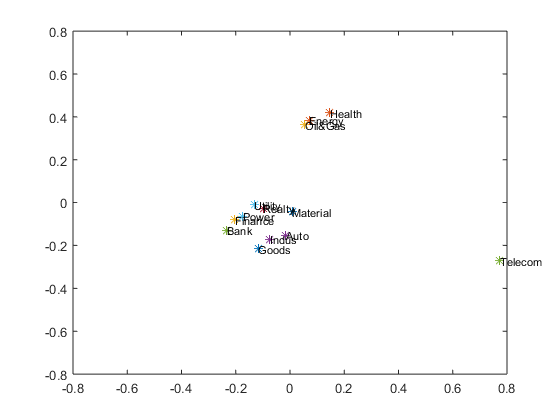}
\caption{Plot of MDS for the indices.}
\label{Fig:MDS_indices}
\end{figure}

\begin{figure}[H]
	\centering
		\includegraphics[width=0.5\linewidth]{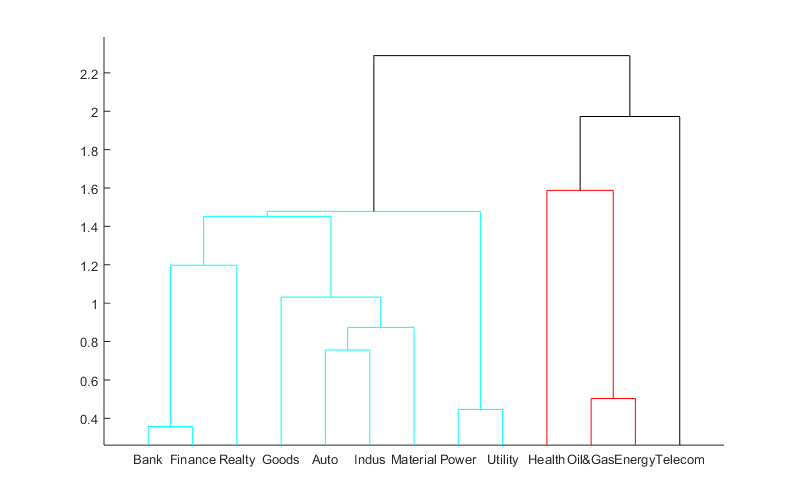}\\
\caption{Plot of dendogram for the indices.}
\label{Fig:dendogram_indices}
\end{figure}

\begin{figure}[H]
	\centering

		\includegraphics[width=0.4\linewidth]{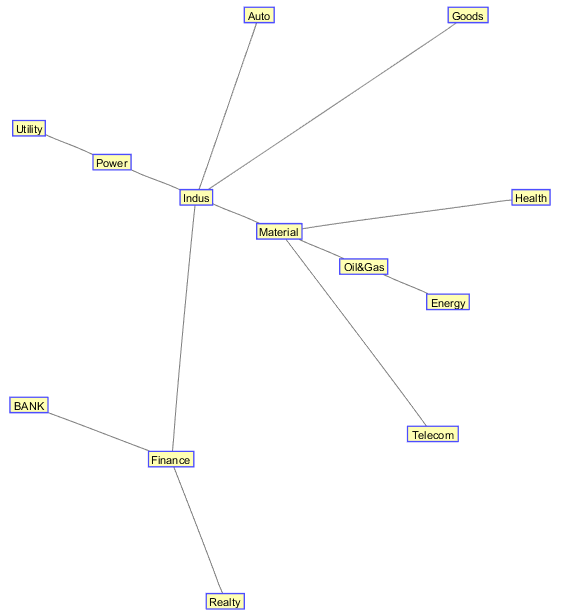}
	\caption{
Plot of MST of BSE Indices.
}
	\label{Fig:Index_plots}
\end{figure}

Empirically we used the 13 sectoral indices from the BSE (list given in appendix II) for the time window May 2015-May 2016.
The resulting multidimensional scaling results have been plotted in Fig. \ref{Fig:MDS_indices},
dendogram in Fig. \ref{Fig:dendogram_indices} as well as 
minimum spanning tree in Fig. \ref{Fig:Index_plots}. 

The MDS algorithm cannot segregate the markets into clusters in a way that corresponds to the industry classifications.
Dendogram produces better results than that. Finally, the minimum spanning tree corresponds to a fairly
intuitive
market structure. Note that the only information used was sectoral returns' correlations. The MST
shows that the banks and realty sectors are most closely related to the finance sector. Energy sector is most
closely associated with oil \& gas sector and so on.
Thus we see that the sectoral MST approximates the industrial relations in a fairly correct manner.


\section{Summary}
In this article we have applied multiple techniques to analyze daily data from Bombay stock exchange. Our analysis covers a large spectrum of tools proposed in the econophysics literature in the last two decades. 
Using eigendecomposition method, we show that the market cross-correlation structure shows a very prominent market mode.
Consistent with the literature, we show that the group mode is not very strong for emerging countries and in fact, is very difficult to differentiate from the random mode. Then we carry out partial correlation analysis, a newly proposed method, on the Indian data.
This helps us to explicitly characterize and quantify the average 'third variable' effect in the cross-correlations.

Finally, we turn to network analysis to study the core-periphery structure. We use multi-dimensional scaling
and dendograms to identify clusters. In general, we do not find any significant pattern between such clusters and the
industrial classifications. However, a much more intuitive picture emerges when we construct 
a mesoscopic network with the sectoral indices. 
We see that minimum spanning tree across the indices clearly segregates nodes according to their industrial classification,
by just using the return cross-correlations.

\vskip 1 cm

\section{Acknowledgement}

This research was partially supported by the institute grant, IIM Ahmedabad.
KS thanks University Grants Commission (Ministry of Human Research Development, Govt. of India)  for her junior research fellowship. AC acknowledges financial support from the institutional research funding IUT (IUT39-1) of the Estonian Ministry of Education and Research, grant number BT/BI/03/004/2003(C) of Govt. of India, Ministry of Science and Technology, Department of Biotechnology, Bioinformatics division, and University of Potential Excellence-II grant (Project ID-47) of the Jawaharlal Nehru University, New Delhi, India.

\section{Appendix I}

\begin{table}[H]
\caption{List of all sectoral indices. The first column has the abbreviation, the second column has the full name of the sector, as given in the BSE.}
	\begin{center}
		\begin{tabular}{|c|c|}
	\hline
	BSESN  &  S\&P BSE SENSEX \\
\hline
SENSEX  &	S\&P BSE SENSEX \\
\hline
BSE500	& S\&P BSE 500 \\
\hline
SI1900	& S\&P BSE AUTO \\
\hline
SIBANK	& S\&P BSE BANKEX \\
\hline
SPBSBMIP  &	S\&P BSE BASIC MATERIALS \\
\hline
SI0200  &	S\&P BSE CAPITAL GOODS \\
\hline
SPBSEIP  &	S\&P BSE ENERGY \\
\hline
SPBSFIIP &	S\&P BSE FINANCE \\
\hline
SPBSIDIP &	S\&P BSE INDUSTRIALS \\
\hline
SI1400  &	 S\&P BSE OIL \& GAS \\
\hline
SIPOWE &	S\&P BSE POWER \\
\hline
SIREAL & S\&P BSE REALTY \\
\hline
SPBSTLIP & S\&P BSE TELECOM\\
\hline
SPBSUTIP &	S\&P BSE UTILITIES \\
\hline
SI0800  & S\&P BSE HEALTHCARE \\
\hline 
\end{tabular}
	\end{center}
\end{table}

\section{Appendix II}

\begin{table}[H]
\caption{List of all stocks considered for the analysis. The first column has the abbreviation, the second column has the full name and the third column specifies the sector as given in the BSE.}
	\begin{center}
		\begin{tabular}{ |p{3 cm}| p{6 cm}| p{6 cm}| }
	\hline
	ABB & ABB INDIA LIMITED & Heavy Electrical Equipment \\
	\hline
	ABIRLANUVO & ADITYA BIRLA NUVO LTD. &	Diversified \\
\hline
 AEGISLOG & 	AEGIS LOGISTICS LTD. &	Oil Marketing and Distribution \\
	\hline
 AMARAJABAT &	AMARA RAJA BATTERIES LTD.	& Auto Parts and Equipment \\
\hline
 AMBALALSA &	AMBALAL SARABHAI ENTERPRISES LTD.	& Pharmaceuticals \\
\hline
 ANDHRAPET &	ANDHRA PETROCHEMICALS LTD. &	Commodity Chemicals \\
\hline
 ANSALAPI  &	ANSAL PROPERTIES and INFRASTRUCTURE LTD. &	Realty \\
\hline
 APPLEFIN  &	APPLE FINANCE LTD. &	Finance (including NBFCs) \\
\hline
 ARVIND	& ARVIND LTD. &	Textiles \\
\hline
 ASIANHOTNR	& ASIAN HOTELS (NORTH) LIMITED &	Hotels \\
\hline
 ASSAMCO & 	ASSAM COMPANY (INDIA) LIMITED	& Tea and Coffee \\
\hline
 ATFL &	AGRO TECH FOODS LTD. &	Other Agricultural Products \\
	\hline
 ATUL & ATUL LTD.	& Agrochemicals \\
	\hline
 ATVPR   &  	ATV PROJECTS INDIA LTD.	& Construction and Engineering \\
	\hline
 AUTOLITIND &	AUTOLITE (INDIA) LTD.	& Auto Parts and Equipment \\
	\hline
 AUTORIDFIN	& AUTORIDERS FINANCE LTD.	& Finance (including NBFCs) \\
	\hline
 BAJAJELEC & 	BAJAJ ELECTRICALS LTD.	& Household Appliances \\
	\hline
 BAJAJHIND &	BAJAJ HINDUSTHAN SUGAR LIMITED	& Sugar \\
	\hline
 BAJFINANCE	& BAJAJ FINANCE LIMITED	& Finance (including NBFCs) \\
	\hline
 BALLARPUR &	BALLARPUR INDUSTRIES LTD.	& Paper and Paper Products \\
	\hline
 BALRAMCHIN	& BALRAMPUR CHINI MILLS LTD. &	Sugar \\
	\hline
 BANARISUG &	BANNARI AMMAN SUGARS LTD.	& Sugar \\
	\hline
 BANCOINDIA	& BANCO PRODUCTS (INDIA) LTD.	& Auto Parts and Equipment \\
	\hline
 BASF   &  BASF INDIA LTD. &	Specialty Chemicals \\
	\hline
 BATAINDIA &	BATA INDIA LTD.	& Footwear \\
	\hline
 
\end{tabular}
	\end{center}
\end{table}

\begin{table}[H]
	\begin{center}
		\begin{tabular}{ |p{3 cm}| p{6 cm}| p{6 cm}| }
		\hline 
		BEL &	BHARAT ELECTRONICS LTD.	& Defence \\
	\hline
 BEML	& BEML LTD.	& Commercial Vehicles \\
	\hline
 BEPL  &	BHANSALI ENGINEERING POLYMERS LTD.	& Specialty Chemicals \\
	\hline
 BHAGGAS &  BHAGAWATI GAS LIMITED	 & Industrial Gases \\
	\hline
 BHEL  &	BHARAT HEAVY ELECTRICALS LTD.	& Heavy Electrical Equipment \\
	\hline
 BHUSANSTL &	BHUSHAN STEEL LTD.& Iron and Steel/Interm.Products \\
\hline 
BIHSPONG  &	BIHAR SPONGE IRON LTD.	& Iron and Steel/Interm.Products \\
\hline
 BINANIIND &	BINANI INDUSTRIES LTD.	& Holding Companies \\
\hline
 BIRLACORPN	& BIRLA CORPORATION LIMITED	& Flagship company \\
\hline 
BIRLAERIC &	BIRLA ERICSSON OPTICAL LTD.	& Other Elect.Equip./ Prod. \\
\hline
 BLUESTARCO	& BLUE STAR LTD.	& Consumer Electronics \\
\hline
 BNKCAP	& BNK CAPITAL MARKETS LTD.	& Other Financial Services \\
\hline
 BOMDYEING &	BOMBAY DYEING and MFG.CO.LTD.	& Textiles \\
\hline
 BPL  & BPL LTD.	& Consumer Electronics \\
\hline
CAMPHOR  &	CAMPHOR and ALLIED PRODUCTS LTD.	& Commodity Chemicals \\
\hline  
 CENTENKA  &	CENTURY ENKA LTD.	& Textiles \\
\hline
 CENTEXT  & 	CENTURY EXTRUSIONS LTD.	& Aluminium \\
\hline
 CENTURYTEX	 & CENTURY TEXTILES and INDUSTRIES LTD.	& Cement and Cement Products \\
\hline
 CESC	& CESC LTD.	& Electric Utilities \\
\hline
CHAMBLFERT	& CHAMBAL FERTILISERS and CHEMICALS LTD.	& Fertilizers \\
\hline
 CHENNPETRO	& CHENNAI PETROLEUM CORPORATION LTD.	& Refineries/ Petro-Products \\
\hline
 CIPLA  & CIPLA LTD.	& Pharmaceuticals \\
\hline
 CMIFPE & CMI FPE LTD. & Industrial Machinery \\
\hline
 CRISIL & CRISIL LTD.	& Other Financial Services \\
\hline
 CROMPGREAV	& CROMPTON GREAVES LTD.	& Heavy Electrical Equipment \\
\hline
 DABUR  & DABUR INDIA LTD.	& Personal Products \\
\hline 
DALMIASUG &	DALMIA BHARAT SUGAR AND INDUSTRIES LTD	& Sugar \\
\hline
DCW  & DCW LTD.	& Petrochemicals \\
\hline
 DHAMPURSUG	& DHAMPUR SUGAR MILLS LTD.	& Sugar \\

\hline
\end{tabular}
	\end{center}
\end{table}

\begin{table}[H]
	\begin{center}
		\begin{tabular}{ |p{3 cm}| p{6 cm}| p{6 cm}| }
		\hline
 DIAMINESQ &	DIAMINES and CHEMICALS LTD.	& Commodity Chemicals \\
\hline
 DICIND  &	DIC INDIA LTD.	& Specialty Chemicals \\
\hline
 DISAQ &	DISA INDIA LTD.	& Industrial Machinery \\
\hline 
DRREDDY  & 	DR.REDDY'S LABORATORIES LTD.	& Pharmaceuticals \\
\hline
 EIDPARRY & E.I.D.-PARRY (INDIA) LTD.	& Sugar \\
\hline
 ELANTAS &  ELANTAS BECK INDIA LTD.	& Commodity Chemicals \\
\hline 
ELECTCAST &	ELECTROSTEEL CASTINGS LTD.	& Construction and Engineering \\
\hline
 EMPEESUG & 	EMPEE SUGARS and CHEMICALS LTD.	& Sugar \\
\hline
 ENVAIREL & ENVAIR ELECTRODYNE LTD.	& Industrial Machinery \\
\hline 
ESABINDIA &	ESAB INDIA LTD.	& Other Industrial Goods \\
\hline
 ESSELPRO & 	ESSEL PROPACK LTD.	& Containers and Packaging \\
\hline 
ESTER  &	ESTER INDUSTRIES LTD.	& Commodity Chemicals \\
\hline
 EXIDEIND &	EXIDE INDUSTRIES LTD.	& Auto Parts and Equipment \\
\hline
 FEDDERLOYD	& FEDDERS LLOYD CORPORATION LTD. & Other Elect.Equip./ Prod. \\
\hline
 FERROALL  &	FERRO ALLOYS CORPORATION LTD.	& Iron and Steel/Interm.Products \\
\hline
 FGP &	FGP LTD.	& Finance (including NBFCs) \\
\hline
 FINCABLES &	FINOLEX CABLES LTD.	& Other Elect.Equip./ Prod. \\
\hline 
FORCEMOT & 	FORCE MOTORS LTD.	& Cars and Utility Vehicles \\
\hline
 FOSECOIND &	FOSECO INDIA LTD.	& Commodity Chemicals \\
\hline
 GANESHBE & GANESH BENZOPLAST LTD.	& Commodity Chemicals \\
\hline
 GARDENSILK	& GARDEN SILK MILLS LTD.	& Textiles \\
\hline 
GHCL  & 	GHCL LTD.	& Commodity Chemicals \\
\hline
 GLFL & 	GUJARAT LEASE FINANCING LTD.	& Finance (including NBFCs) \\
\hline
 GODFRYPHLP	& GODFREY PHILLIPS INDIA LTD.	& Cigarettes-Tobacco Products \\
\hline
GODREJIND 	& GODREJ INDUSTRIES LTD.	& Commodity Chemicals \\
		\hline 
 GOLDENTOBC	& GOLDEN TOBACCO LTD.	& Cigarettes-Tobacco Products \\
\hline
 GOODRICKE &	GOODRICKE GROUP LTD.	& Tea and Coffee \\
\hline
 GOODYEAR & 	GOODYEAR INDIA LTD.	& Auto Tyres and Rubber Products \\
\hline
 GRASIM  &  GRASIM INDUSTRIES LTD.	& Textiles \\
\hline 
		GTL  & 	GTL LTD.	& Other Telecom Services \\
\hline
 GTNINDS &	GTN INDUSTRIES LTD.	& Textiles \\

\hline
\end{tabular}
	\end{center}
\end{table}

\begin{table}[H]
	\begin{center}
		\begin{tabular}{ |p{3 cm}| p{6 cm}| p{6 cm}| }
		\hline
 GUJFLUORO &	GUJARAT FLUOROCHEMICALS LTD.	& Industrial Gases \\
\hline
 HCC   &	HINDUSTAN CONSTRUCTION CO.LTD.	& Construction and Engineering \\
\hline
 HCIL  &	HIMADRI CHEMICALS and INDUSTRIES LTD. &	Commodity Chemicals \\
\hline
HDFC  &	HOUSING DEVELOPMENT FINANCE CORP.LTD.	& Housing Finance \\
\hline
 HDFCBANK &	HDFC BANK LTD &	Banks \\
\hline 
HEIDELBERG	& HEIDELBERGCEMENT INDIA LTD.	& Cement and Cement Products \\
\hline
 HEROMOTOCO &	HERO MOTOCORP LTD.	& $2/3$ Wheelers \\
\hline
 HFCL  & HIMACHAL FUTURISTIC COMMUNICATIONS LTD.	& Telecom Cables \\
\hline
 HINDOILEXP	& HINDUSTAN OIL EXPLORATION CO.LTD.	& Exploration and Production \\
\hline 
HINDPETRO &	HINDUSTAN PETROLEUM CORPORATION LTD.	& Refineries/ Petro-Products \\
\hline
 HINDUJAVEN	& HINDUJA VENTURES LTD.	& Broadcasting and Cable TV \\
\hline
 HINDZINC & 	HINDUSTAN ZINC LTD.	& Zinc \\
\hline 
HMT  & HMT LTD.	& Commercial Vehicles \\
\hline
HOTELEELA &	HOTEL LEELAVENTURE LTD.	& Hotels \\
\hline 
HSIL  & HSIL LTD.	& Containers and Packaging \\
\hline
IDBI  & IDBI BANK LTD.	& Banks \\
\hline
IFCI  & IFCI LTD.	& Financial Institutions \\
\hline 
IFSL  & INTEGRATED FINANCIAL SERVICES LTD.	& Finance (including NBFCs) \\
\hline
IGPL  & I G PETROCHEMICALS LTD.	& Commodity Chemicals \\
\hline 
INDIAGLYCO	& INDIA GLYCOLS LTD.	& Commodity Chemicals \\
\hline
INDLEASE  &	INDIA LEASE DEVELOPMENT LTD.	& Finance (including NBFCs) \\
\hline 
INDORAMA  &	INDO RAMA SYNTHETICS (INDIA) LTD.	& Textiles \\
\hline
 INDSUCR  & INDIAN SUCROSE LTD.	& Beverage Store \\
\hline
 INFY  &	INFOSYS LTD.	& IT Consulting and Software \\
\hline 
INGERRAND &	INGERSOLL-RAND (INDIA) LTD.	& Industrial Machinery \\
\hline
 INSILCO  & 	INSILCO LTD.	& Other Industrial Goods \\
\hline
IONEXCHANG	& ION EXCHANGE (INDIA) LTD.	& Industrial Machinery \\

\hline
\end{tabular}
	\end{center}
\end{table}

\begin{table}[H]
	\begin{center}
		\begin{tabular}{ |p{3 cm}| p{6 cm}| p{6 cm}| }
		\hline 
ITHL  & INTERNATIONAL TRAVEL HOUSE LTD.	& Travel Support Services \\
\hline
JASCH  & JASCH INDUSTRIES LTD.	& Textiles \\
\hline
JAYKAY  & JAYKAY ENTERPRISES LTD.	& Finance (including NBFCs) \\
\hline
JCTLTD   & JCT LTD.	& Textiles \\
\hline
JINDALPOLY	& JINDAL POLY FILMS LTD.	& Commodity Chemicals \\
\hline
JISLJALEQS	& JAIN IRRIGATION SYSTEMS LTD.	& Plastic Products \\
		\hline  
JKLAKSHMI &	JK LAKSHMI CEMENT LTD.	& Cement and Cement Products \\
\hline
JSWSTEEL  &	JSW STEEL LTD.	& Iron and Steel/Interm.Products \\
\hline
KAJARIACER &	KAJARIA CERAMICS LTD.	& Furniture-Furnishing-Paints \\
\hline 
KAKATCEM & 	KAKATIYA CEMENT SUGAR and INDUSTRIES LTD.	& Cement and Cement Products \\
\hline
KANELIND  &	KANEL INDUSTRIES LTD.	& Comm.Trading and Distribution \\
\hline
KANSAINER &	KANSAI NEROLAC PAINTS LTD. &	Furniture-Furnishing-Paints \\
\hline
KGDENIM  & KG DENIM LTD.	& Textiles \\
\hline
KINETICENG	& KINETIC ENGINEERING LTD. &	2/3 Wheelers \\
\hline
KIRLFER & KIRLOSKAR FERROUS INDUSTRIES LTD.	& Iron and Steel/Interm.Products \\
\hline 
KIRLOSBROS &	KIRLOSKAR BROTHERS LTD.	& Industrial Machinery \\
\hline
KIRLOSIND &	KIRLOSKAR INDUSTRIES LTD. &	Industrial Machinery \\
 \hline
KOTAKBANK &	KOTAK MAHINDRA BANK LTD.	& Banks \\
\hline
KSBPUMPS  &	KSB PUMPS LTD. &	Industrial Machinery \\
\hline
KSL  & KALYANI STEELS LTD. &	Iron and Steel/Interm.Products \\
\hline
LAXMIMACH &	LAKSHMI MACHINE WORKS LTD. &	Industrial Machinery \\
\hline 
LGBBROSLTD &	L.G.BALAKRISHNAN and BROS.LTD. &	Auto Parts and Equipment \\
\hline
LICHSGFIN &	LIC HOUSING FINANCE LTD. &	Housing Finance \\
\hline 
LML & LML LTD. &	2/3 Wheelers \\
\hline
LOKHSG  & LOK HOUSING and CONSTRUCTIONS LTD. &	Realty \\
\hline
LORDSCHLO & LORDS CHLORO ALKALI LTD. &	Commodity Chemicals \\
\hline
LUPIN  &  LUPIN LTD. &	Pharmaceuticals \\
\hline
LYKALABS & LYKA LABS LTD. &	Pharmaceuticals \\
\hline
MAFATIND & MAFATLAL INDUSTRIES LTD. &	Textiles \\
\hline
MAHSCOOTER &	MAHARASHTRA SCOOTERS LTD. &	2/3 Wheelers \\

\hline

\end{tabular}
	\end{center}
\end{table}		

\begin{table}[H]
	\begin{center}
		\begin{tabular}{ |p{3 cm}| p{6 cm}| p{6 cm}| }
		\hline
MAHSEAMLES	& MAHARASHTRA SEAMLESS LTD. &	Construction and Engineering \\
\hline 
MAJESAUT  &	MAJESTIC AUTO LTD. &	2/3 Wheelers \\
\hline
MANALIPETC &	MANALI PETROCHEMICAL LTD. &	Petrochemicals \\
\hline
MARGOFIN & MARGO FINANCE LTD. &	Finance (including NBFCs) \\
\hline
MAVIIND  & MAVI INDUSTRIES LTD. &	Plastic Products \\
\hline
MERCK  & MERCK LTD. &	Pharmaceuticals \\
\hline
METROGLOBL &	METROGLOBAL LTD.	& Paper and Paper Products \\
\hline 
MFSL  & MAX FINANCIAL SERVICES LTD. &	Life Insurance \\
\hline
MIDINDIA  &	MID INDIA INDUSTRIES LTD. &	Textiles \\
\hline
MIRCELECTR &	MIRC ELECTRONICS LTD. &	Consumer Electronics \\
\hline 
MOREPENLAB	& MOREPEN LABORATORIES LTD. &	Pharmaceuticals \\
\hline
MRF  &  MRF LTD. &	Auto Tyres and Rubber Products \\
\hline
MRPL &  MANGALORE REFINERY and PETROCHEMICALS LTD. &	Refineries/ Petro-Products \\
\hline 
MTNL & MAHANAGAR TELEPHONE NIGAM LTD. &	Telecom Services \\
\hline
NAHARSPING &	NAHAR SPINNING MILLS LTD. &	Textiles \\
		\hline 
NATPEROX  &	NATIONAL PEROXIDE LTD. &	Commodity Chemicals \\
\hline 
NCC  & NCC LTD.	& Construction and Engineering \\
\hline
NEPCMICON &	NEPC INDIA LTD. &	Heavy Electrical Equipment \\
\hline 
NIITLTD  & NIIT LTD. &	IT Training Services \\
\hline
NIRLON  & NIRLON LTD. &	Misc.Commercial Services \\
\hline 
OILCOUNTUB	& OIL COUNTRY TUBULAR LTD.	& Oil \\
\hline
ONGC  & OIL AND NATURAL GAS CORPORATION LTD.  & Oil and Gas \\
\hline 
ORIENTBANK	& ORIENTAL BANK OF COMMERCE	& Banks \\
\hline
ORIENTHOT &	ORIENTAL HOTELS LTD.	& Hotels \\
\hline
OSWALAGRO &	OSWAL AGRO MILLS LTD. &	Real Estate \\
\hline 
PANCM  & PANYAM CEMENTS AND MINERAL INDS. &	Cement and Cement Products \\
\hline
PARRYSUGAR &	PARRYS SUGAR INDUSTRIES LTD. &	Sugar \\
\hline
PDUMJEPULP &	PUDUMJEE PULP AND PAPER MILLS LTD.	& Paper and Paper Products \\
\hline

\end{tabular}
	\end{center}
\end{table}

\begin{table}[H]
	\begin{center}
		\begin{tabular}{ |p{3 cm}| p{6 cm}| p{6 cm}| }
		\hline 
		PEL &  PIRAMAL ENTERPRISES LTD. &	Pharmaceuticals \\
		\hline
	PENTAGRAPH &	PENTAMEDIA GRAPHICS LTD.	& Graphics \\
\hline
PHCAP  & PH CAPITAL LTD. &	Comm.Trading and Distribution \\
\hline
PIDILITIND	& PIDILITE INDUSTRIES LTD.	& Manufacturer \\
\hline
PILITA  &	PIL ITALICA LIFESTYLE LTD.	& Plastic Furniture Manufacturers \\
\hline
PIXTRANS & PIX TRANSMISSIONS LTD. &	Manufacturer of Industrial Belts, Automotive Belts and Agricultural Belts \\
\hline
PRAGBOS & PRAG BOSIMI SYNTHETICS LTD. &	Textiles \\
\hline
PRIMESECU & 	PRIME SECURITIES LTD. &	Security \\
\hline
PRISMCEM & PRISM CEMENT LTD.?	& Cement and Cement Products \\
\hline
PUNJCOMMU &	PUNJAB COMMUNICATIONS LTD.	& Communication \\
\hline
RAIN  & RAIN INDUSTRIES LTD.	& Rain Cements Limited \\
\hline
RAJSREESUG	& RAJSHREE SUGAR AND CHEMICAL LTD. &	Sugar \\
\hline 
RALLIS  & RALLIS INDIA LIMITED- NITYA AGRO SERVICES	& Agrochemicals Supplier \\
\hline
RAMANEWS & SHREE RAMA NEWSPRINT LTD.	& Newsprint and Papers \\
\hline
RAMCOCEM & THE RAMCO CEMENTS LTD. &	Cement and Cement Products \\
\hline 
RAYMOND & RAYMOND GROUP	& Fabrics \\
\hline
RELCAPITAL &	RELIANCE CAPITAL LTD. &	Finance (including NBFCs) \\
\hline
RSWM  & R S W M Ltd. &	Textiles \\
\hline
SAIL  & STEEL AUTHORITY OF INDIA LTD. &	Iron and Steel/Interm.Products \\
\hline
SBI   &  STATE BANK OF INDIA &	Banks \\
\hline
SBIN  &	STATE BANK OF INDIA &	Banks \\
\hline
SPICEJET  &	SPICEJET LTD. &	Airlines \\
\hline 
SURYAROSNI &	SURYA ROSHNI LTD. &	Misc.Commercial Services \\
\hline
TITAN  &  TITAN COMPANY LTD.	& Other Apparels and Accessories \\
\hline 
TRENT  & TRENT LTD. &	Department Stores \\
\hline
UFLEX  & UFLEX LTD. &	Containers and Packaging \\
\hline
UMANGDAIR &	UMANG DAIRIES LTD. &	Packaged Foods \\
\hline 
VEDL & VEDANTA LTD. &	Iron and Steel/Interm.Products \\
\hline
WHIRLPOOL &	WHIRLPOOL OF INDIA LTD. &	Consumer Electronics \\
	\hline

\end{tabular}
	\end{center}
\end{table}


\bibliographystyle{plain}

\end{document}